\documentclass[aps,reprint,superscriptaddress,citeautoscript,floatfix,longbibliography,bibnotes,graphicx]{revtex4-2}
\usepackage{times,siunitx,amsmath,amsfonts,amsthm,amssymb,amsbsy,braket,graphicx,bbm,hyperref,orcidlink}

\usepackage[utf8]{inputenc}
\usepackage[T1]{fontenc}
\usepackage[capitalize]{cleveref}
\DeclareFontFamily{OT1}{pzc}{}
\DeclareFontShape{OT1}{pzc}{m}{it}{<-> s * [1.10] pzcmi7t}{}
\DeclareMathAlphabet{\mathpzc}{OT1}{pzc}{m}{it}
\newcommand{\up}{{\uparrow}}
\newcommand{\down}{{\downarrow}}
\newcommand{\updown}{{\uparrow\downarrow}}
\newcommand{\id}{\mathbbm 1}
\newcommand{\ii}{\mathrm{i}} 
\newcommand{\dd}{\mathrm{d}} 
\newcommand{\ee}{\mathrm{e}}

\renewcommand{\Re}{\mathrm{Re}}
\renewcommand{\Im}{\mathrm{Im}}
\DeclareMathOperator{\sgn}{sgn}
\DeclareMathOperator{\pf}{pf}

\hypersetup{colorlinks = true, urlcolor=blue, linkcolor=blue, citecolor = red}
 
\begin{document}
\title[Majorana/Andreev crossover and topological phase transition in inhomogeneous nanowires]{
Majorana/Andreev crossover and the fate of the topological phase transition in inhomogeneous nanowires
}
\author{Pasquale Marra\orcidlink{0000-0002-9545-3314}}
\email{pmarra@ms.u-tokyo.ac.jp}
\address{
Graduate School of Mathematical Sciences,
The University of Tokyo, 3-8-1 Komaba, Meguro, Tokyo, 153-8914, Japan}
\address{
Department of Physics, and Research and Education Center for Natural Sciences, 
Keio University, 4-1-1 Hiyoshi, Yokohama, Kanagawa, 223-8521, Japan}

\author{Angela Nigro\orcidlink{0000-0001-8326-5781}}
\address{
Dipartimento di Fisica ``E. R. Caianiello'', Università degli Studi di Salerno, 84084 Fisciano (Salerno), Italy}
\address{
Consiglio Nazionale delle Ricerche CNR-SPIN, UOS Salerno, 84084 Fisciano (Salerno), Italy}

\date{\today}

\begin{abstract}
Majorana bound states (MBS) and Andreev bound states (ABS) in realistic Majorana nanowires setups have similar experimental signatures which make them hard to distinguishing one from the other. Here, we characterize the continuous Majorana/Andreev crossover interpolating between fully-separated, partially-separated, and fully-overlapping Majorana modes, in terms of global and local topological invariants, fermion parity, quasiparticle densities, Majorana pseudospin and spin polarizations, density overlaps and transition probabilities between opposite Majorana components. We found that in inhomogeneous wires, the transition between fully-overlapping trivial ABS and nontrivial MBS does not necessarily mandate the closing of the bulk gap of quasiparticle excitations, but a simple parity crossing of partially-separated Majorana modes (ps-MM) from trivial to nontrivial regimes. We demonstrate that fully-separated and fully-overlapping Majorana modes correspond to the two limiting cases at the opposite sides of a continuous crossover: the only distinction between the two can be obtained by estimating the degree of separations of the Majorana components. This result does not contradict the bulk-edge correspondence: Indeed, the field inhomogeneities driving the Majorana/Andreev crossover have a length scale comparable with the nanowire length, and therefore correspond to a nonlocal perturbation which breaks the topological protection of the MBS\@.
\end{abstract}
\maketitle

\section{Introduction}

Majorana bound states (MBS) can emerge as topologically protected and spatially-separated zero-energy excitations localized at the opposite ends of a one-dimensional (1D) topological superconductor~\cite{kitaev_unpaired_2001,oreg_helical_2010,lutchyn_majorana_2010,alicea_new_2012,leijnse_introduction_2012,stanescu_majorana_2013,beenakker_search_2013,elliott_colloquium_2015,beenakker_randommatrix_2015,sato_majorana_2016,sato_topological_2017,aguado_majorana_2017,lutchyn_majorana_2018,zhang_next_2019,li_exploring_2019,frolov_topological_2020,flensberg_engineered_2021}.
Their nonabelian exchange statistics~\cite{kitaev_unpaired_2001,ivanov_non-abelian_2001,kitaev_fault-tolerant_2003}
may lead to the realization of fault-tolerant quantum computation~\cite{nayak_nonabelian_2008,pachos_introduction_2012,das-sarma_majorana_2015,lahtinen_short_2017,stanescu_introduction_2017,beenakker_search_2020,oreg_majorana_2020,aguado_perspective_2020}.
1D topological superconductivity can be realized in Majorana nanowires, i.e., proximitized semiconducting nanowires with strong spin-orbit coupling and broken time-reversal symmetry~\cite{alicea_new_2012,leijnse_introduction_2012,stanescu_majorana_2013,beenakker_search_2013,elliott_colloquium_2015,beenakker_randommatrix_2015,sato_majorana_2016,sato_topological_2017,aguado_majorana_2017,lutchyn_majorana_2018,zhang_next_2019,li_exploring_2019,frolov_topological_2020,flensberg_engineered_2021},
in epitaxial 1D semiconductor-superconductor heterostructures~\cite{shabani_two-dimensional_2016,hell_two-dimensional_2017,pientka_topological_2017},
arrays of magnetic atoms deposited on a conventional superconductor~\cite{choy_majorana_2011,nadj-perge_proposal_2013,pientka_topological_2013,braunecker_interplay_2013,klinovaja_topological_2013,vazifeh_self-organized_2013,li_topological_2014,kim_helical_2014,pientka_unconventional_2014,heimes_majorana_2014,brydon_topological_2015},
or optically-trapped ultracold fermionic atoms coupled to a molecular BEC cloud~\cite{jiang_majorana_2011,nascimbene_realizing_2013,buhler_majorana_2014,ptok_quantum_2018}.

Quite a few experiments observed signatures compatible with the existence of MBSs in Majorana nanowires, e.g., zero-bias conductance peaks (ZBCP) in extended regions of the phase diagram~\cite{mourik_signatures_2012,das_zero-bias_2012,deng_anomalous_2012,churchill_superconductor-nanowire_2013,finck_anomalous_2013,deng_majorana_2016,nichele_scaling_2017,chen_experimental_2017,deng_nonlocality_2018,gul_ballistic_2018,bommer_spin-orbit_2019,zhang_large_2021} 
and in multi-terminal devices~\cite{grivnin_concomitant_2019,puglia_closing_2021,heedt_shadow-wall_2021},
fractional Josephson effect~\cite{rokhinson_the-fractional_2012,laroche_observation_2019, dartiailh_phase_2021}, 
and Coulomb blockade spectroscopy~\cite{higginbotham_parity_2015,albrecht_exponential_2016,albrecht_transport_2017,shen_parity_2018,vaitiekenas_flux-induced_2020}.
However, similar signatures may as well be induced by Andreev bound states (ABS) with zero or near-zero energies appearing in the topologically trivial phase~\cite{de-moor_electric_2018,chen_ubiquitous_2019,anselmetti_end-to-end_2019,menard_conductance-matrix_2020,junger_magnetic-field-independent_2020,yu_non-majorana_2021,valentini_nontopological_2021}. 

There are several physical mechanisms leading to the emergence of these trivial ABS in Majorana nanowires~\cite{prada_andreev_2020}:
random disorder~\cite{liu_zero-bias_2012,pikulin_a-zero-voltage_2012,bagrets_class_2012,roy_topologically_2013,stanescu_disentangling_2013,pan_physical_2020,pan_generic_2020,das-sarma_disorder-induced_2021,pan_disorder_2021,pan_crossover_2021,pan_three-terminal_2021,pan_quantized_2021},
localized impurities~\cite{stanescu_nonlocality_2014,pan_crossover_2021}, 
strong interband coupling~\cite{bagrets_class_2012,woods_zero-energy_2019},
and spatial inhomogeneities induced by smooth potentials~\cite{kells_near-zero-energy_2012,prada_transport_2012,stanescu_nonlocality_2014,liu_andreev_2017,penaranda_quantifying_2018,moore_two-terminal_2018,avila_non-hermitian_2019,vuik_reproducing_2019,stanescu_robust_2019,woods_zero-energy_2019,pan_physical_2020,pan_crossover_2021,pan_quantized_2021,pan_three-terminal_2021},
quantum dots~\cite{kells_near-zero-energy_2012,prada_transport_2012,stanescu_disentangling_2013,liu_andreev_2017,setiawan_electron_2017,ptok_controlling_2017,penaranda_quantifying_2018,moore_quantized_2018,moore_two-terminal_2018,liu_distinguishing_2018,reeg_zero-energy_2018,huang_metamorphosis_2018,stanescu_robust_2019,avila_non-hermitian_2019,vuik_reproducing_2019,lai_presence_2019,sharma_hybridization_2020,pan_physical_2020,pan_three-terminal_2021,pan_quantized_2021,hess_local_2021},
or partial proximization~\cite{kells_near-zero-energy_2012,prada_transport_2012,chevallier_mutation_2012,stanescu_disentangling_2013,cayao_sns-junctions_2015,liu_andreev_2017,setiawan_electron_2017,penaranda_quantifying_2018,moore_quantized_2018,moore_two-terminal_2018,liu_distinguishing_2018,reeg_zero-energy_2018,huang_metamorphosis_2018,fleckenstein_decaying_2018,avila_non-hermitian_2019,vuik_reproducing_2019,stanescu_robust_2019,lai_presence_2019,sharma_hybridization_2020,pan_physical_2020,pan_three-terminal_2021,hess_local_2021,liu_majorana_2021}.
Inhomogeneous potentials naturally arise also in ultracold-atom setups due to the presence of the optical-trap used to confine the atomic cloud~\cite{jiang_majorana_2011,nascimbene_realizing_2013,buhler_majorana_2014,ptok_quantum_2018}.
Indeed, trivial ABS may exhibit quantized ZBCP virtually indistinguishable from those produced by MBS~\cite{liu_andreev_2017,moore_quantized_2018,moore_two-terminal_2018,liu_distinguishing_2018,lai_presence_2019,pan_physical_2020,pan_crossover_2021,pan_generic_2020,das-sarma_disorder-induced_2021,pan_quantized_2021,hess_local_2021}, 
mimic the oscillations of the energy splitting of MBS in some regimes~\cite{cao_decays_2019,sharma_hybridization_2020}, 
be robust against local perturbations~\cite{penaranda_quantifying_2018,stanescu_robust_2019}, 
and even exhibit nonabelian braiding statistics~\cite{vuik_reproducing_2019,zeng_feasibility_2020}.
However, whereas MBS are exponentially localized at the edges of the nanowire or, equivalently, at a topological domain wall, ABS are localized anywhere inside the wire, typically near inhomogeneities or impurities, and do not necessarily exhibit exponential localization~\cite{penaranda_quantifying_2018,moore_two-terminal_2018,avila_non-hermitian_2019,stanescu_robust_2019,pan_quantized_2021}.

In a single-band, infinitely long, clean and homogenous Majorana nanowire (i.e., with uniform chemical potential $\mu$, Zeeman field $b$, and superconducting pairing $\Delta$), the presence/absence of MBS correspond to the realization of the nontrivial/trivial topological phase, separated by a sharp topological quantum phase transition (TQPT)~\cite{lutchyn_majorana_2010,oreg_helical_2010}.
MBS are described by a nonlocal fermionic state as a superposition of two fully-separated Majorana modes exponentially localized at the opposite edges of the wire, with zero overlap and zero energy, topologically robust against local perturbations, and exhibit an exactly quantized ZBCP $G=2e^2/h$ at both wire ends.
This case is well-described as a "black and white" dichotomy:
the wire is either topologically nontrivial $|b|>|b_c|\equiv\sqrt{\mu^2+\Delta^2}$, 
with the presence of perfectly self-conjugate, exponentially-localized, and fully-separated Majorana modes with \emph{exactly} zero-energy, 
or topologically trivial $|b|<|b_c|$, 
with no energy states below the bulk gap~\cite{lutchyn_majorana_2010,oreg_helical_2010}.
The two phases are separated by a well-defined TQPT which coincides with the closing of the bulk gap $\Delta E$.
We note that, even in clean and homogenous wires, MBS are a limiting case, having zero-energy only for $L\to\infty$:
In finite wires $L<\infty$, Majorana modes hybridize and gain a finite energy~\cite{prada_transport_2012,klinovaja_composite_2012,das-sarma_splitting_2012,rainis_towards_2013,stanescu_dimensional_2013,fleckenstein_decaying_2018}.

Unfortunately, in the laboratory everything is finite, and spatial inhomogeneities, disorder, and impurities are practically unavoidable.
These effects produce a rich variety of physical regimes which cannot be described in terms of a "black and white" dichotomy.
For instance, the properties of ABS induced by random disorder or localized impurities differ qualitatively from ABS induced by spatial inhomogeneities.
Disorder-induced and impurity-induced ABS have small or nearly-zero energy, are localized away from the wire edges, and with ZBCP $G\gtrsim2e^2/h$ in an extended window of the parameter space, typically without end-to-end correlations~\cite{pikulin_a-zero-voltage_2012,stanescu_disentangling_2013,stanescu_nonlocality_2014,lai_presence_2019,pan_physical_2020,pan_generic_2020,pan_quantized_2021,pan_crossover_2021}.
They may appear in the topologically trivial phase, even at zero magnetic field.
By increasing the magnetic field up to the critical value $b_c$, the system may or may not reach the nontrivial phase at intermediate or strong disorder regimes~\cite{pan_crossover_2021}.
Inhomogeneity-induced ABS instead exhibit small energies oscillating in the magnetic field, are localized near the inhomogeneities, with a nearly-quantized ZBCP which may appear only at one end of the wire~\cite{moore_quantized_2018,moore_two-terminal_2018,penaranda_quantifying_2018,avila_non-hermitian_2019,stanescu_robust_2019}.
They always appear at finite fields, below or above the critical field $b_c$, and exhibit properties that are somewhat similar to those of topologically nontrivial MBS\@.

The distinction between MBS and ABS can be understood in terms of spatial separation of the Majorana components.
It is a well-known fact that any fermionic mode $d$ can be decomposed as a sum of two Majorana modes $d=\gamma_A+\ii\gamma_B$.
The case where the two modes $\gamma_A$ and $\gamma_B$ are localized at the opposite edges of topologically nontrivial wire of infinite length correspond to topologically protected MBS\@.
On the other hand, the case where the two modes $\gamma_A$ and $\gamma_B$ are fully or partially overlapping correspond to subgap ABS or overlapping MBS in short wires.
The mutual overlap between the Majorana modes and their susceptibility against local perturbations can be quantified by spatial integrals~\cite{penaranda_quantifying_2018}.
It is known that wavefunctions of inhomogeneity-induced ABS show a partial-separation of the Majorana components, and can be connected to topologically nontrivial MBS via a continuous crossover from the trivial $b<b_c$ to the nontrivial regime $b>b_c$~\cite{moore_two-terminal_2018,stanescu_robust_2019,avila_non-hermitian_2019}.
Moreover, it has been shown recently that there exists a crossover between inhomogeneity-induced ABS and disorder-induced ABS~\cite{pan_crossover_2021}.
The presence of these two distinct crossovers may seem paradoxical: 
inhomogeneity-induced ABS can be continuously transformed into nontrivial MBS or trivial ABS\@.
On top of that, these crossovers may occur with or without a sharp quantum phase transition~\cite{prada_transport_2012,stanescu_to-close_2012,stanescu_disentangling_2013,mishmash_approaching_2016,huang_metamorphosis_2018}
where trivial ABS detach from the bulk excitation spectrum and gradually approach zero-energy as the magnetic fields increases~\cite{prada_transport_2012,penaranda_quantifying_2018,moore_two-terminal_2018,vuik_reproducing_2019}.
Therefore, it is natural to ask, is there a continuous crossover between topologically trivial and topologically nontrivial MBS\@?
If this crossover occurs, what is the fate of the topological phase transition?

To answer these questions, we consider single-band clean Majorana nanowires of finite length $L$ described by the Oreg-Lutchyn model~\cite{lutchyn_majorana_2010,oreg_helical_2010} in the presence of spatial inhomogeneities and impurities, 
and characterize the crossover between topologically trivial impurity-induced ABS, inhomogeneities-induced ABS, and topologically nontrivial MBS\@.
Considering several different potential landscapes, we disentangle the different physical regimes separately in terms of topology, fermion parity crossings, and localization properties of the subgap states.
In particular, we define and calculate the global and local topological invariants, local Majorana mass, and the fermion parity, and classify the different regimes of the Majorana wire into three different phases, i.e., 
the homogeneous topologically trivial phase	(TTP),
the homogeneous topologically nontrivial phase (TNP),
and a topologically inhomogeneous phase (TIP) separating the first two phases.
We then characterize the localization and mutual overlap of Majorana modes via the
quasiparticle densities, Majorana pseudospin and spin polarizations, density overlaps and transition probabilities.
This lead us to distinguish between 
fully-separated Majorana modes (fs-MM),
fully-overlapping Majorana modes (fo-MM),
and
partially-separated Majorana modes (ps-MM) which interpolate between the first two cases.
These different kinds of MM can be characterized by the mutual overlap $\Omega$ and transition probability $W$ between their two Majorana components, and expectation values of the Majorana pseudospin.

We find that the Majorana/Andreev crossover 
from impurity-induced ABS to quasi-MBS~\cite{pan_crossover_2021} and
from inhomogeneities-induced ABS (quasi-MBS) to MBS~\cite{moore_two-terminal_2018,stanescu_robust_2019,avila_non-hermitian_2019}
can be described as a transition between the two limiting cases of fs-MM and fo-MM, which can be alternatively viewed as a fusion of two MM into a single Dirac-fermion mode. 
The MM localize at points where the local Majorana mass becomes close to zero, at the nodes of the local Majorana mass, or at the edges of the wire. 
The first case is realized in the TTP, where the local Majorana mass is always positive, and fo-MM localize at the minima of the local Majorana mass. 
The second case is realized in the TIP, where ps-MM localize at the nodes of the local Majorana mass.
These ps-MM can become fs-MM if the distance between the Majorana modes approaches infinity.
The last case is realized in the TNP, where ps-MM localize at the edges of the wire. 
These ps-MM become fs-MM in the limit of infinitely long wire.
The transition between these different phases do not necessarily correspond to the condition $|b|\equiv\sqrt{\mu^2+\Delta^2}$, and does not necessarily coincide with the closing and reopening of the bulk gap, but only to the presence of fermion parity crossings of the lowest energy (LE) subgap state.
For these reasons, we argue that the ps-MM in the TIP and TTP are indistinguishable from a physical point of view.

\section{Methods}

\subsection{Hamiltonian}

We consider a Majorana nanowire, i.e., a semiconducting nanowire with strong spin-orbit coupling (e.g., InAs, InSb) in a magnetic field, and proximitized superconductivity induced by a conventional $s$-wave superconductor (e.g. Al), described by the Oreg-Lutchyn model 
\begin{align}\label{eq:TheContinuousHamiltonian}
H= \left( \frac{p^2}{2m} + \frac{\alpha}{\hbar} \sigma_y p - 
\mu(x) 
+ {b}(x) \cdot {\sigma}_z \right) \tau_z
+ \Delta(x) \ii\sigma_y \tau_x,
\end{align}
where
$p=-\ii\hbar\partial_x$ is the momentum operator, 
$\sigma_i$ and
$\tau_i$ with $i=x,y,z$
the Pauli matrices in spin space and particle-hole space, 
$m$ the effective mass of the wire, $\alpha$ the spin-orbit coupling strength, $\mu(x)$ the chemical potential, 
$b(x)=({g}/2)\mu_B {B}(x)$ the Zeeman field in the $z$-direction (perpendicular to the spin-orbit coupling), and $\Delta(x)\ge0$ the superconducting pairing due to proximization.
A more realistic nanowire model can be obtained via a self-energy term describing the proximitized superconductivity in the wire~\cite{stanescu_proximity_2010,stanescu_proximity-induced_2017} and solving the resulting Hamiltonian self-consistently.
However, self-energy corrections may be neglected since they do not affect the quasiparticle states near zero energy~\cite{pan_crossover_2021}.
Moreover, in ultracold-atom setups, the superconducting term depends on the quasiparticle densities, and in that case the corresponding Bogoliubov-de~Gennes equations needs to be solved self-consistently~\cite{jiang_majorana_2011,nascimbene_realizing_2013,buhler_majorana_2014,ptok_quantum_2018}.

Notice that the Hamiltonian is real, and consequently its eigenstates have real wavefunctions (up to a global phase).
The Hamiltonian is indeed in the BDI symmetry class~\cite{kitaev_periodic_2009,schnyder_classification_2009,ryu_topological_2010,chiu_classification_2016} with unbroken particle-hole symmetry $\mathcal C=\tau_x \mathcal K$ and unbroken ``time-reversal'' symmetry $\mathcal T'=\mathcal K$, where $\mathcal K$ is the complex conjugate operator.
We consider a InSb/Al Majorana wire of length $L={2000}~\mathrm{nm}$, $m=0.015\ m_e\approx {7700}~\mathrm{eV}/c^2$, $\alpha={0.5}~{\mathrm{eV}\AA}$,
$b/B={1.5}~\mathrm{meV/T}$ (i.e., $g$-factor $g\approx50$), 
$\Delta(x)=\Delta={1}~\mathrm{meV}$ (see Ref.~\onlinecite{lutchyn_majorana_2018}).
We discretize the continuous Hamiltonian into a tight-binding Hamiltonian via finite-difference method on a discrete lattice with lattice constant $a={10}~\mathrm{nm}$ and calculate the energy spectra and the density of states (DOS) numerically via exact diagonalization.
The code used for the numerical calculations can be found on Zenodo~\cite{marra_data_2021}.

\subsection{Quasiparticle densities }


To determine the localization of Majorana modes along the wire, we discretize the Hamiltonian in \cref{eq:TheContinuousHamiltonian} and diagonalize it to obtain the full energy spectra.
In particular we focus on the subgap energy level and the corresponding wavefunctions, described in terms of the Nambu bi-spinor
\begin{equation}\label{eq:fermion}
\psi(x)=\begin{pmatrix} u_{\up}(x)\\ u_{\down}(x)\\ v_{\up}(x)\\ v_{\down}(x) \end{pmatrix}.
\end{equation}
We then calculate the quasiparticle density $\rho(x)$ as the sum of the densities of the particle and hole sectors, i.e.,
\begin{equation}\label{eq:psi}
|\psi(x)|^2= \sum_{\sigma=\updown} |u_{\sigma}(x)|^2+ |v_{\sigma}(x)|^2.
\end{equation}

To disentangle the Majorana/Andreev nature of subgap states, we need to look to the quasiparticle density 
in the Majorana basis.
Any fermionic state $c$ can be decomposed into two Majorana modes 
$c=(\gamma_A+\ii\gamma_B)/\sqrt{2}$ with $\gamma_A=(c^\dag+c)/\sqrt{2}$ and $\gamma_B=(c^\dag - c)/\sqrt{2}\ii$. 
Hence, the fermionic states in \cref{eq:fermion} can be decomposed into two Majorana components (compare with, e.g., Ref.~\cite{penaranda_quantifying_2018}) given by
\begin{subequations}
\begin{align}
\psi_A(x)=
\frac1{\sqrt2}(\id+\mathcal C)\psi(x)=
\frac1{\sqrt2}
&
\begin{pmatrix}
u_{\up}(x)+\overline v_{\up}(x)
\\ 
u_{\down}(x)+\overline v_{\down}(x)
\\
v_{\up}(x)+\overline u_{\up}(x)
\\ 
v_{\down}(x)+\overline u_{\down}(x)
\end{pmatrix}
,
\\
\psi_B(x)=
\frac1{\ii\sqrt2}(\id-\mathcal C)\psi(x)=
\frac1{\ii\sqrt2}
&
\begin{pmatrix}
u_{\up}(x)-\overline v_{\up}(x)
\\ 
u_{\down}(x)-\overline v_{\down}(x)
\\
v_{\up}(x)-\overline u_{\up}(x)
\\ 
v_{\down}(x)-\overline u_{\down}(x)
\end{pmatrix}
.
\end{align}
\end{subequations}
We then calculate the densities of the two Majorana components as
\begin{subequations}\label{eq:psiAB}
\begin{align}
|\psi_A(x)|^2=& \sum_{\sigma=\updown} 
|u_{\sigma}(x) + \overline v_{\sigma}(x)|^2,
\\
|\psi_B(x)|^2=& \sum_{\sigma=\updown} 
|u_{\sigma}(x) - \overline v_{\sigma}(x)|^2,
\end{align}
\end{subequations}
which can be thought as the ``partial'' quasiparticle density with respect to the two Majorana flavors A and B, with $|\psi(x)|^2=(|\psi_A(x)|^2+|\psi_B(x)|^2)/2$.

For fs-MM, we expect two density peaks localized at the opposite ends of the wire, with the two Majorana components fully separated, i.e., with
$|\psi_A(x)|^2=|\psi(x)|^2$ and $|\psi_B(x)|^2=0$ on one edge, and 
$|\psi_B(x)|^2=|\psi(x)|^2$ and $|\psi_A(x)|^2=0$ on the opposite edge.
On the other hand, for fo-MM, we naturally expect 
$|\psi_A(x)|^2=|\psi_B(x)|^2=|\psi(x)|^2$, i.e., the two components of the Majorana modes are fully overlapping.

\subsection{Majorana pseudospin}

\begin{figure*}[t]
\centering
\includegraphics[width=\textwidth]{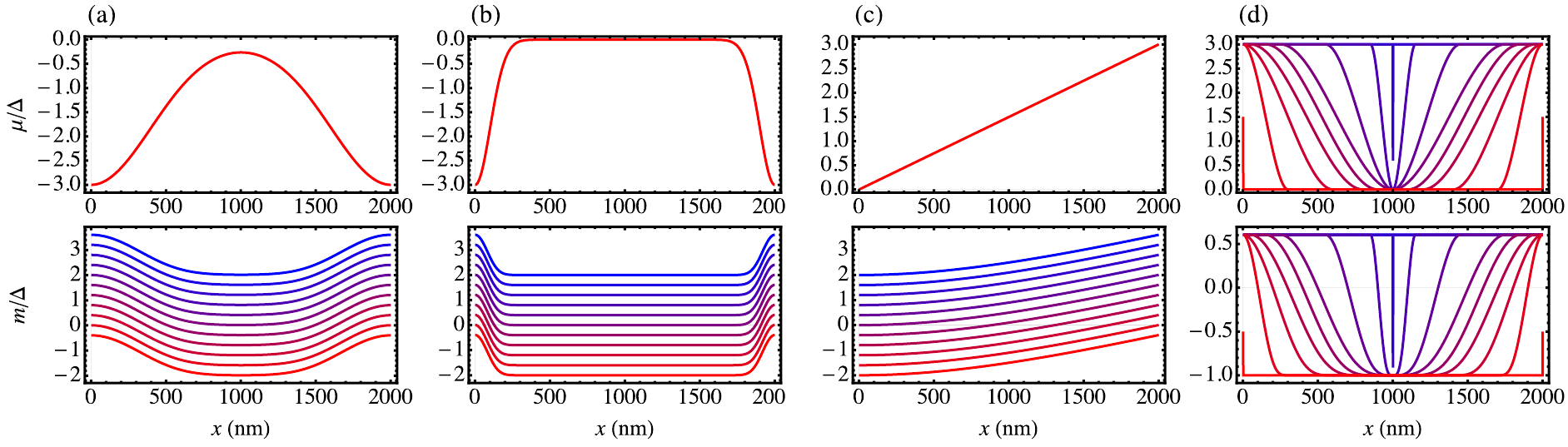} 
\caption{
Spatial dependence of the chemical potential and resulting Majorana mass for different shapes of the potential.
(a) and (b) smooth potential barriers in \cref{eq:smoothbarriers} respectively for $w=0.2L$ (a) and $w=0.05L$ for different choices of the magnetic field $0\le b\le 2\Delta$,
(c) linear slope in \cref{eq:potentialslope} for different choices of the magnetic field $0\le b\le 2\Delta$,
and
(d) smoothly interpolating potential defined in \cref{eq:smoothinterpolation,eq:interpolation} for different choices of the control parameter $0\le r\le 4$.
}
\label{fig:shapes}
\end{figure*}


Following Refs.~\onlinecite{marra_1d-majorana_2021,marra_dispersive_2021} we define the Majorana pseudospin operator $\mathbf T=\boldsymbol\tau/2$ as the analogous of the spin operator $\mathbf S=\boldsymbol\sigma/2$.
The expectation values of the Majorana pseudospin are given by $\braket{\mathbf T(x)}=\bra{\psi(x)}\mathbf T \ket{\psi(x)}$ with cartesian components given by
\begin{subequations}
\begin{align}
\braket{T_x(x)}=&
\sum_{\sigma=\updown} 
\Re \left( \overline u_{\sigma}(x) v_{\sigma}(x) \right),
\\
\braket{T_y(x)}=&
\sum_{\sigma=\updown} 
\Im \left( \overline u_{\sigma}(x) v_{\sigma}(x)\right),
\\
\braket{T_z(x)}=&
\sum_{\sigma=\updown} 
\frac12\left(
|u_{\sigma}(x)|^2 -|v_{\sigma}(x)|^2\right),
\end{align}
\end{subequations}
The expectation values of the $x,y$ components of the Majorana pseudospin coincide with the Majorana polarization introduced in Ref.~\onlinecite{sticlet_spin_2012}.
Notice also that, since the Hamiltonian in \cref{eq:TheContinuousHamiltonian} and its eigenstates are real, one has $\braket{T_y(x)}=0$.
From \cref{eq:psi,eq:psiAB} it follows that
\begin{subequations}\label{eq:psiAB}
\begin{align}
|\psi_A(x)|^2=& |\psi(x)|^2+2\langle T_x(x)\rangle,
\\
|\psi_B(x)|^2=& |\psi(x)|^2-2\langle T_x(x)\rangle,
\end{align}
\end{subequations}

For fs-MM, we expect $\braket{T_z(x)}=0$ and $\braket{T_x(x)}\neq0$ at the edges of the wire, with $\braket{T_x(x)}>0$ and $\braket{T_x(x)}<0$ at opposite edges.
Conversely, for fo-MM, we expect $\braket{T_x(x)}=0$ and $\braket{T_z(x)}>0$ for particle-like excitations, and $\braket{T_z(x)}<0$ for hole-like excitations.
Hence, Dirac-fermion excitations (i.e., non Majorana) have Majorana pseudospin on the $z$ axis, while
Majorana-like excitations have pseudospin perpendicular to the $z$ axis.
Since for fs-MM one has that $\braket{T_x(x)}>0$ and $\braket{T_x(x)}<0$ at opposite edges, the average pseudospin along the wire is zero.
However, the average of the square of the pseudospin, defined as
\begin{equation}
\braket{T_{i}^2}=
\int \dd x \braket{T_{i}(x)}^2
\end{equation}
with $i=x,y,z$, is expected to be nonzero for a fs-MM.

For comparison, we also calculate the expectation values of the spin $\braket{\mathbf S(x)}=\bra{\psi(x)}\mathbf S \ket{\psi(x)}$ with cartesian components given by
\begin{subequations}
\begin{align}
\braket{S_x(x)}=&
\sum_{w=u,v} 
\Re \left( \overline w_{\up}(x) w_{\down}(x) \right),
\\
\braket{S_y(x)}=&
\sum_{w=u,v} 
\Im \left( \overline w_{\up}(x) w_{\down}(x)\right),
\\
\braket{S_z(x)}=&
\sum_{w=u,v} 
\frac12\left(
|w_{\up}(x)|^2 -|w_{\down}(x)|^2\right).
\end{align}
\end{subequations}
To characterize the overall spin polarization of subgap modes, we consider the quantities
\begin{equation}
\braket{S_{i}^2}=
\int \dd x \braket{S_{i}(x)}^2,
\end{equation}
with $i=x,y,z$.


\subsection{Overlaps and matrix elements}

We can characterize the mutual overlaps of the Majorana modes in terms of 
the overlaps of the quasiparticle densities and 
the matrix elements of the Hamiltonian between the Majorana components.
Assuming a single subgap energy level, the overlaps between quasiparticle densities of the Majorana components are 
\begin{equation}
\Omega=
\int\dd x\,
|\psi_A(x)|
|\psi_B(x)|,
\end{equation}
introduced in Ref.~\onlinecite{penaranda_quantifying_2018}.
Moreover, the transition probabilities between the two Majorana modes are given by their matrix element
\begin{equation}
W=-\ii
\int\dd x\,
\overline\psi_A(x)
H\psi_B(x).
\end{equation}
The matrix elements allows one to construct the effective Hamiltonian
\begin{equation}\label{eq:Heff}
H_\text{eff}=\ii W \gamma_A\gamma_B =\frac12
\begin{pmatrix}\gamma_A&\gamma_B\end{pmatrix}
\begin{pmatrix}
0 		& \ii W	\\
- \ii W		& 0	\\
\end{pmatrix}
\begin{pmatrix}\gamma_A\\ \gamma_B\end{pmatrix},
\end{equation}
which describes the low-energy spectra of the Majorana wire in the limit where the energy of the subgap modes are small compared with the bulk gap $W/\Delta E\approx0$.
For fs-MM, there is no overlap of the quasiparticle densities, which mandates $\Omega=W/\Delta E=0$.
For fo-MM instead, one has $|\psi_A(x)|=|\psi_B(x)|$ which gives $\Omega=1$ and $W/\Delta E\approx1$.

\subsection{Global topological invariant and fermion parity}

\begin{figure*}[t]
\centering
\includegraphics[width=\textwidth]{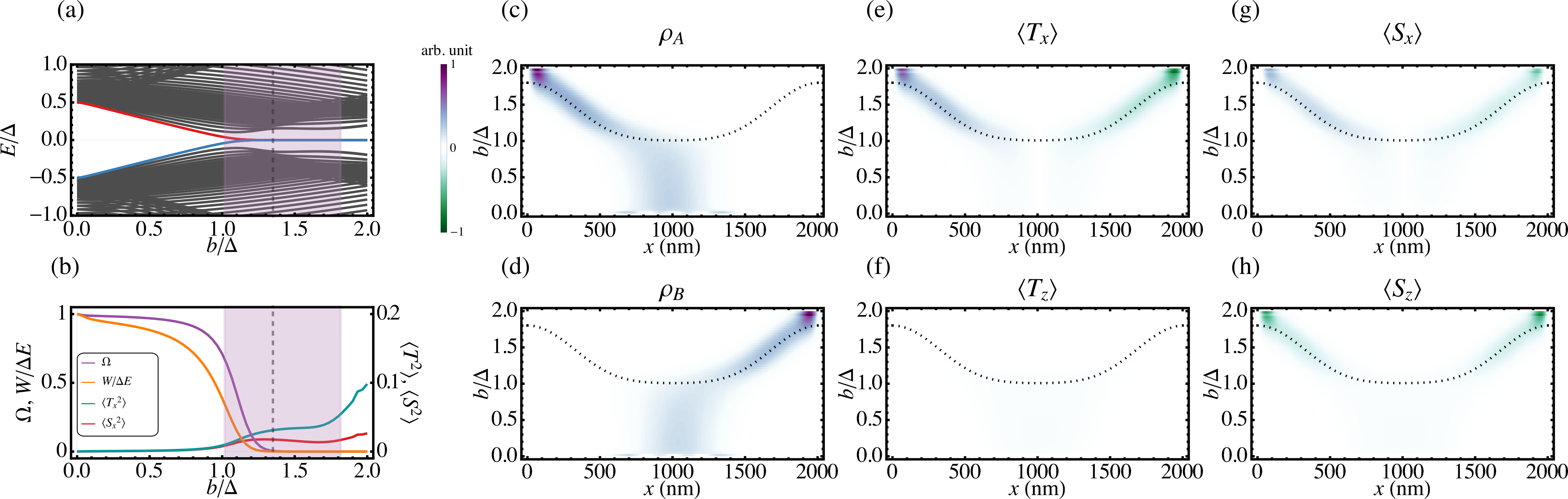} 
\caption{
Numerical results for a Majorana wire with OBC and smooth potential barriers with $w=0.2L$, as a function of the Zeeman field.
(a) Energy spectrum with the two particle-hole symmetric lowest energy (LE) levels highlighted in color.
The shaded area corresponds to the topologically inhomogeneous phase (TIP) separating the topologically trivial phase (TTP) and the topologically nontrivial phase (TNP).
The vertical line indicates the fermion parity crossing of the groundstate with periodic boundary conditions (PBC).
The LE level corresponds to two Majorana modes (MM) which are either fully overlapping (fo-MM) at zero field, or partially-separated (ps-MM).
(b) The mutual overlap $\Omega$ and transition probability $W$ between the two components of the MM, compared with the expectation values of the square of the Majorana pseudospin $\braket{T_x^2}$ and spin $\braket{S_x^2}$ integrated along the whole wire.
(c), (d) Quasiparticle densities of the two components of the MM as a function of the position, which show a clear spatial separation in the TIP and TNP\@.
(e), (f) Expectation values of the Majorana pseudospin components $\braket{T_x(x)}$ and $\braket{T_z(x)}$ respectively.
(g), (h) Expectation values of the spin components $\braket{S_x(x)}$ and $\braket{S_z(x)}$ respectively.
The peaks of the quasiparticle densities, Majorana pseudospin, and spin, are localized near the nodes of the local Majorana mass (dotted line) in the TIP\@. 
The numerical data plotted here and in the following figures can be found on Zenodo~\cite{marra_data_2021}.
}
\label{fig:gaussian1a}
\end{figure*}

\begin{figure*}[t]
\centering
\includegraphics[width=\textwidth]{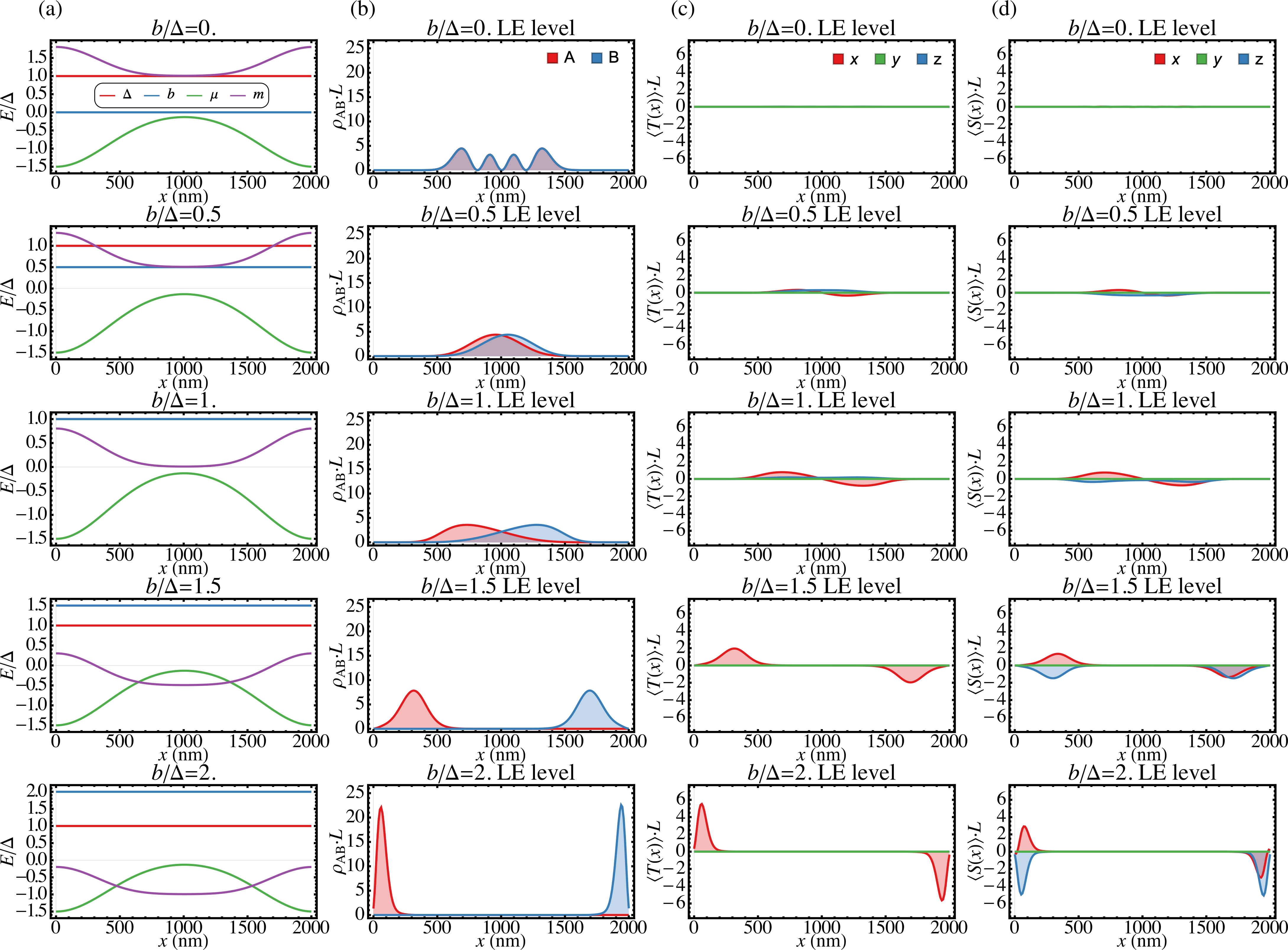} 
\caption{
Snapshots of the numerical results of \cref{fig:gaussian1a} at several different Zeeman fields.
(a) Chemical potential, Zeeman field, and superconducting pairing, and Majorana mass along the wire.
(b) Quasiparticle densities of the two components of the MM.
(c), (d) The three components of the Majorana pseudospin and of the spin of the MM.
}
\label{fig:gaussian1b}
\end{figure*}


In the case of uniform fields, the $\mathbb{Z}_2$ topological invariant of the Majorana wire is $\mathcal P=\sgn \mathcal M$, where $\mathcal M=\sqrt{\mu^2+\Delta^2}-|b|$ is the Majorana mass.
Topologically trivial and nontrivial phases are realized respectively for $\mathcal M>0$ and $\mathcal M<0$~\cite{oreg_helical_2010,lutchyn_majorana_2010}.
We recall that the lowest energy sector of the Oreg-Lutchyn minimal model~\cite{oreg_helical_2010,lutchyn_majorana_2010} is unitarily equivalent to a Dirac equation in the Majorana representation and with a Majorana mass equal to $\mathcal M$ (see Ref.~\onlinecite[page 198-202]{shen_topological_2017}).

In the presence of small inhomogeneities or weak disorder the $\mathbb{Z}_2$ invariant can be generalized~\cite{budich_equivalent_2013} 
as the fermion parity of the groundstate of the system with periodic boundary conditions (PBC).
The fermion parity is equal to
\begin{equation}\label{eq:GlobalTI}
\mathcal P=
\sgn\,\pf\left(
\ii \mathcal H\tau_x
\right),
\end{equation}
where the $\mathcal H$ is the matrix obtained by discretizing the continuous Hamiltonian $H$ on a discrete lattice.
We will calculate the Pfaffian in \cref{eq:GlobalTI} numerically~\cite{wimmer_algorithm_2012}.

For a finite wire with ps-MM with a finite overlap $W>0$, the fermion parity of the PBC groundstate is also equal to the sign of the Pfaffian~\cite{kitaev_unpaired_2001} of the effective Hamiltonian in \cref{eq:Heff} rewritten in the Majorana basis, which yields
\begin{equation}\label{eq:FPW}
{\mathcal P}\equiv{\mathcal P}_\text{eff}=\sgn\pf 
\begin{pmatrix}
0 		& W	\\
- W		& 0	\\
\end{pmatrix}
=\sgn W.
\end{equation}
As a consequence, the transition between trivial and nontrivial topological phase coincide with the fermion parity crossing of the ps-MM (calculated for PBC).

\subsection{Local topological invariant}

\begin{figure*}[t]
\centering
\includegraphics[width=\textwidth]{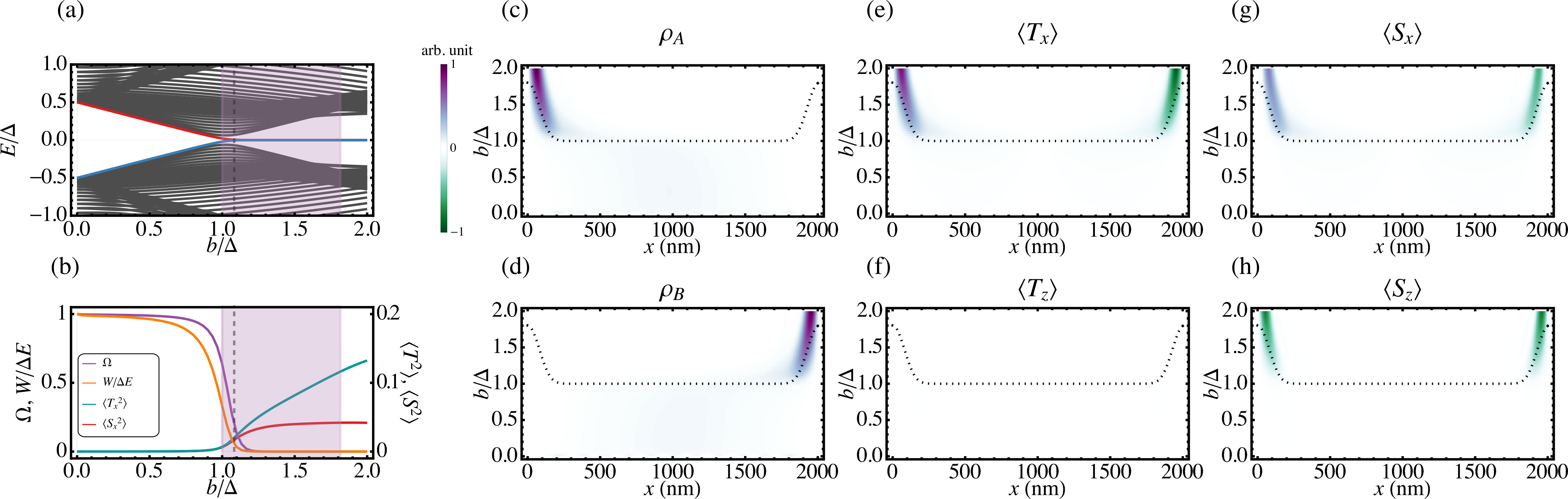} 
\caption{
Numerical results for a Majorana wire with OBC and smooth potential barriers with $w=0.05L$, as a function of the Zeeman field.
}
\label{fig:gaussian2a}
\end{figure*}

\begin{figure*}[t]
\centering
\includegraphics[width=\textwidth]{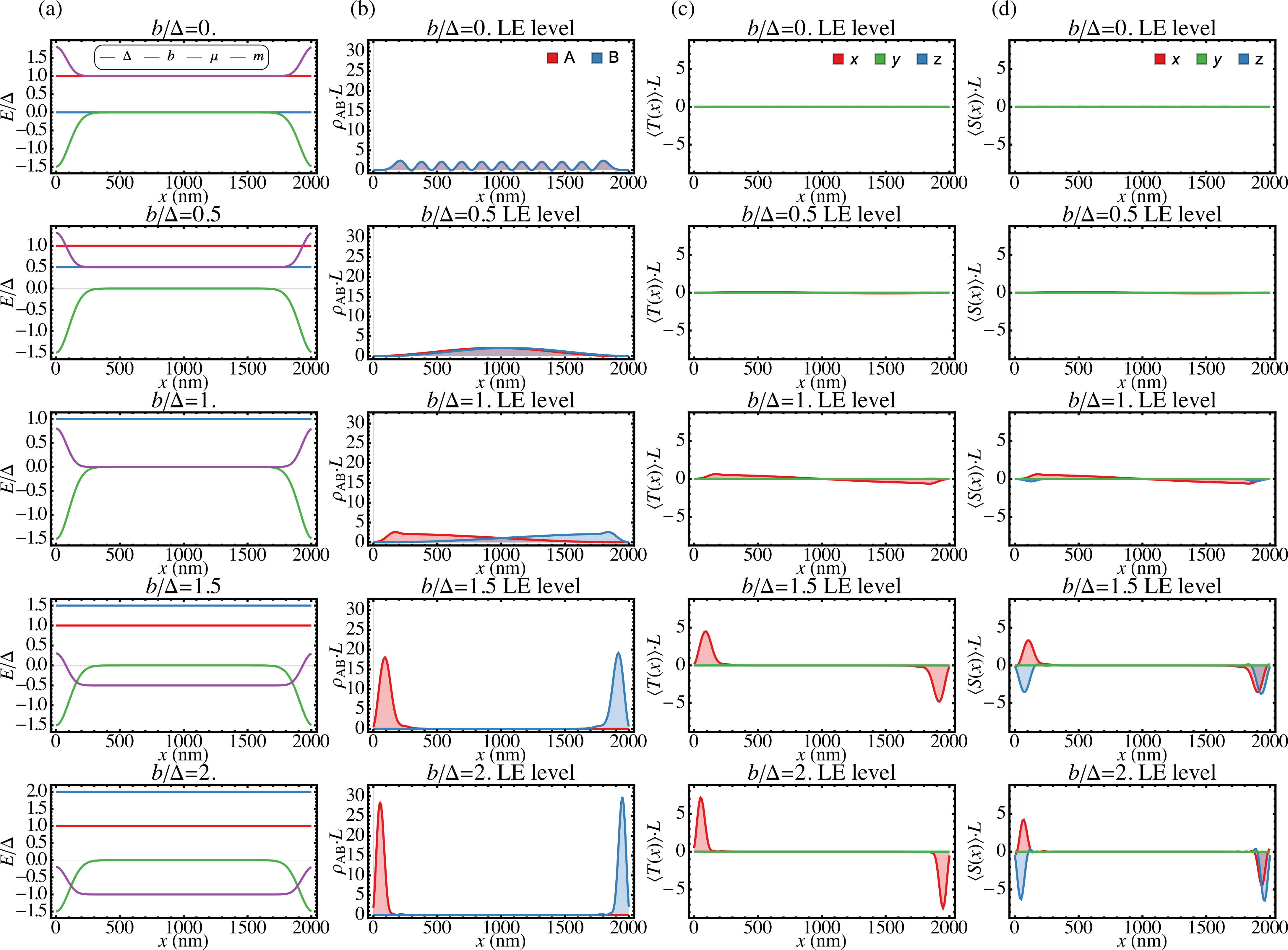} 
\caption{
Snapshots of the numerical results of \cref{fig:gaussian2a} at several different Zeeman fields.
}
\label{fig:gaussian2b}
\end{figure*}

In the presence of inhomogeneous fields, it makes sense to define a \emph{local} Majorana mass and a \emph{local} topological invariant as
\begin{subequations}
\begin{align}
 \mathpzc{m}(x)&=\sqrt{\mu(x)^2+\Delta(x)^2}-|b(x)|,
\\
 \mathpzc{p}(x)&=\sgn{\mathpzc{m}(x)}.
\end{align}
\end{subequations}
The local topological invariant of an inhomogeneous system $\mathpzc{p}(x)$ can be defined in general as the topological invariant of a homogeneous system where all fields are uniform and equal to the values of the fields at the point $x$.
It can be equivalently defined as the topological invariant of a subsystem~\cite{grabsch_pfaffian_2019} which coincide with the infinitesimal segment $[x,x+\dd x]$.
If the spatial variations of the fields are small enough, the local topological invariant is constant along the wire and coincides with the global topological invariant (fermion parity) $\mathpzc{p}(x)\equiv\mathcal P$ in \cref{eq:GlobalTI}.

However, if spatial variations are large, the local topological invariant may be not constant.
In particular, the local Majorana mass may assume alternatively positive and negative values: Segments with $\mathpzc{m}(x)>0$ and $\mathpzc{m}(x)<0$ realize topologically trivial and nontrivial phases with $\mathpzc{p}(x)=\pm1$ respectively.
Hence, we distinguish three phases:
the homogeneous topologically trivial phase	(TTP) where $\mathpzc{m}(x)>0$ along the whole wire,
the homogeneous topologically nontrivial phase (TNP) where $\mathpzc{m}(x)<0$ along the whole wire,
and the topologically inhomogeneous phase (TIP) where $\mathpzc{m}(x)$ changes its sign along the whole wire.
A simple criterium to distinguish the homogeneous and inhomogeneous phases is
\begin{equation}
\sgn\left(\min(\mathpzc{m}(x))\max(\mathpzc{m}(x))\right)=
\begin{cases}
+1\Rightarrow\text{TTP or TNP}\\
-1\Rightarrow\text{TIP}
\end{cases}
\end{equation}
The fermion parity of the PBC groundstate is even and odd respectively in the TTP and TNP\@.
Consequently, on the intermediate TIP, the subgap levels must exhibit an odd number (at least one) of fermion parity crossings of the PBC groundstate.

In the TIP, assuming that the lengths of the trivial and nontrivial segments are larger than the Majorana localization length $\xi_\text{M}$, 
we expect that Majorana modes localize at the boundaries between trivial and nontrivial segments, i.e., at the nodes of the local Majorana mass $\mathpzc{m}(x)=0$.
If the transition between positive and negative mass $\mathpzc{m}(x)$ is sufficiently sharp, one may expect that the localization length of the Majorana modes close to the nominal Majorana localization length $\xi_\text{M}$, which is $\xi\approx\alpha/\Delta$ or $\xi\approx(b/E_\text{SO})\alpha/\Delta$ respectively for strong and weak spin-orbit coupling regimes $E_\text{SO}=m\alpha^2/2\hbar^2\gg\Delta$ and $\ll\Delta$~\cite{klinovaja_composite_2012,mishmash_approaching_2016,aguado_majorana_2017}.

\subsection{Fermion parity and topological invariant}

One may be tempted to identify the fermion parity of the PBC groundstate as the topological invariant even in the TIP\@.
However, this may not be consistent with the bulk-edge correspondence.
To illustrate this, one can consider a simple counterexample of a wire with 
$\mu=0$ near the edges $x<L/2$ and $x>3L/4$ and
$\mu=\Delta$ near the center $L/4<x<3L/4$.
The Majorana mass is consequently 
$\mathpzc{m}(x)=\Delta-b$ near the edges and 
$\mathpzc{m}(x)=\sqrt{2}\Delta-b$ near the center.
For $b<\Delta$ the wire is in the TTP with fermion parity $\mathcal{P}=1$,
for $b>\sqrt{2}\Delta$ it is in the TNP with fermion parity $\mathcal{P}=-1$, 
whereas for $\Delta<b<\sqrt{2}\Delta$, the wire is in the TIP\@.
Thus, the Majorana mass is
$\mathpzc{m}(x)<0$ for $x<L/2$ and $x>3L/4$ (near the edges) and $\mathpzc{m}(x)>0$ for $L/4<x<3L/4$ (near the center).
By increasing the magnetic field $b$ within the TIP, there will necessarily exist a point $b^*<\sqrt{2}\Delta$ where the fermion parity changes its sign, so that $\mathcal{P}=-1$ for $b>b^*$.
For values $b^*<b<\sqrt{2}\Delta$, the wire is in the TIP, with the Majorana mass changing sign along the wire, and with an odd fermion parity of the PBC\@.
Now, if we impose open boundary conditions (OBC) at $x=0$ and $x=L$, the wire develops two MM localized at the edges, since $\mathpzc{m}(x)<0$ near the edges.
Then one can translate the whole wire by $x\to x+L/2$ using PBC\@.
In this second, unitarily equivalent configuration, the wire has still fermion parity $\mathcal{P}=-1$ for $b>b^*$, but there are no MM at the edges, since now one has $\mathpzc{m}(x)>0$ near the edges.
Therefore, in the TIP, the existence of MM at the edges does not depend on the fermion parity of the PBC groundstate, but only on the value of the local topological invariant near the edges of the wire.

\section{Results}


\subsection{Crossover via smooth confinement}

Potential fluctuation at the end of the nanowire may create inhomogeneous potential barriers.
The precise spatial dependence of the resulting potential is usually not known, but can be reasonably approximated as a generic Gaussian-shaped function~\cite{moore_two-terminal_2018,vuik_reproducing_2019,stanescu_robust_2019,pan_physical_2020,zeng_feasibility_2020,pan_quantized_2021}.
We thus consider smooth potential barriers at both edges of the wire, given by
\begin{equation}\label{eq:smoothbarriers}
\mu(x)=-V(x)=-
\delta V \left(
\ee^{
-\frac{x^2}{2w^2}
}
+
\ee^{
-\frac{(x-L)^2}{2w^2}
}
\right),
\end{equation}
with amplitude
$\delta V =1.5\Delta$
and width $w=0.05L$ or $w=0.2L$,
and keep the Zeeman field and the superconducting pairing uniform along the wire $b(x)=b$, $\Delta(x)=\Delta$. 
In this configuration we calculate the energy spectra, fermion parity, and other physical quantities as a function of the magnetic field, for a wire $L={2000}~\mathrm{nm}$ with OBC\@.
For $b<\Delta$, the wire is in the TTP and the Majorana mass is $\mathpzc{m}(x)>0$ along the whole wire.
For $\Delta<b<\sqrt{\delta V^2+\Delta^2}$, the wire is in the TIP with the Majorana mass $\mathpzc{m}(x)>0$ in the central section and $\mathpzc{m}(x)<0$ near the edges, with the nodes of the Majorana mass given by the solutions of the equation $\sqrt{V(x)^2+\Delta^2}\equiv b$.
For $b>\sqrt{\delta V^2+\Delta^2}$, the wire reaches the TNP and the Majorana mass is $\mathpzc{m}(x)<0$ along the whole wire.
The spatial dependence of the chemical potential in \cref{eq:smoothbarriers} and the resulting Majorana mass for different choices of the magnetic field are shown in \cref{fig:shapes}(a) and (b) respectively for $w=0.2L$ and $w=0.05L$.

\Cref{fig:gaussian1a} shows the energy spectra, the 
mutual overlap, transition probability, quasiparticle densities, expectation values of the Majorana pseudospin and spin of the MM, calculated for a Majorana wire with OBC and smooth potential barriers with $w=0.2L$, as a function of the Zeeman field.
\Cref{fig:gaussian1b} show snapshots of the quasiparticle densities, expectation values of the Majorana pseudospin and spin of the MM, for different values of the Zeeman field.
As the Zeeman field increases, the wire goes from the TTP to the TNP passing by the TIP [shaded areas of \cref{fig:gaussian1a}(a) and (b)].
A fo-MM detaches from the bulk excitation spectra in TTP and transmutes into 
a ps-MM at low energy localized at the nodes of the local Majorana mass in the TIP\@.
By increasing the Zeeman field the two MM move continuously from the central region of the wire to the edges, following the nodes of the Majorana mass in the TIP\@.
The peaks in the density, Majorana pseudospin and spin at the nodes have a Gaussian shape, i.e., the MM are smoothly localized at the nodes of the Majorana mass.
When the Majorana mass become negative along the whole wire in the TNP, the ps-MM are maximally separated at the two opposite ends of the wire, and become exponentially localized.
The crossover from bulk fo-MM into the maximally separated ps-MM occurs in the TIP, with the concurrent change of the fermion parity of the PBC groundstate from $\mathcal P=1$ (TTP and TIP at lower fields) to $\mathcal P=-1$ (TNP and TIP at higher fields).
This crossover is well captured by the mutual overlaps, transition probabilities, and Majorana pseudospin and spin [see \cref{fig:gaussian1a}(b)].
The mutual overlap, transition probability are $\Omega,W/\Delta E\approx1$ for the fo-MM and ps-MM in the TTP, and quickly decays reaching $\Omega,W/\Delta E\approx0$ already in the TIP at higher Zeeman fields.
Conversely, the Majorana pseudospin and spin increase in the TIP\@.
The two components of the ps-MM are separated in space [see \cref{fig:gaussian1a}(c) and (d)] and the expectation values of the Majorana pseudospin and spin along the $x$ axis have opposite values [see \cref{fig:gaussian1a}(e) and (g)].

\Cref{fig:gaussian2a,fig:gaussian2b} show the same as before, but smooth potential barriers with $w=0.05L$.
The peaks of the quasiparticle density localized at the nodes of the Majorana mass are much sharper in the case where the width of the Gaussian barrier $w$ is smaller, whereas the peaks are smoothened out in the case of larger $w$.
This indicates a spatial broadening of the ps-MM in the presence of smooth variations of the Majorana mass, and corresponds to a larger overlap between the two components of the ps-MM.

Another relevant feature in the case $w=0.05L\approx0$ is that the appearance of the ps-MM at low energy is accompanied by an apparent closing and reopening of the bulk gap [see \cref{fig:gaussian2a}(a)].
The gap closing correspond at $b=\Delta$ correspond to the presence of a MM localized in an extended region of the wire [see \cref{fig:gaussian2b}(b)].
However, the closing of the bulk gap is exact only in the limit where the width of the Gaussian barriers vanishes $w\to0$ which correspond to the limit case of a pristine homogeneous nanowire.
Indeed, in the previous case $w=0.2L$ the bulk gap does not completely close [see \cref{fig:gaussian1a}(a)], and the regime $b=\Delta$ correspond to the presence of a MM localized in a narrow region of the wire [see \cref{fig:gaussian1b}(b)].

\subsection{Crossover via linear slope}

\begin{figure*}[t]
\centering
\includegraphics[width=\textwidth]{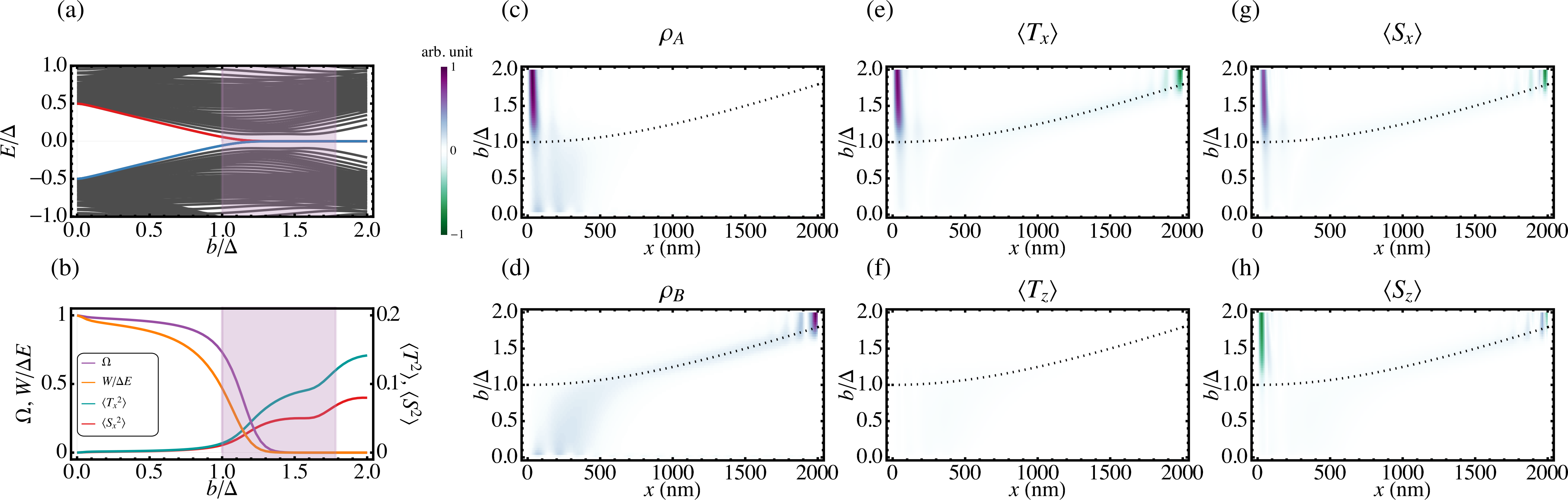} 
\caption{
Numerical results for a Majorana wire with OBC and linear potential slope as a function of the Zeeman field.
}
\label{fig:slopea}
\end{figure*}

We consider a linear variation of the potential (linear slope)~\cite{moore_two-terminal_2018,stanescu_robust_2019}, given by
\begin{equation}\label{eq:potentialslope}
\mu(x)=-V(x)=
\delta V x/L
,
\end{equation}
with amplitude
$\delta V =1.5\Delta$, and keep the Zeeman field and the superconducting pairing uniform along the wire $b(x)=b$, $\Delta(x)=\Delta$.
As before, we calculate the energy spectra, fermion parity, and other physical quantities as a function of the magnetic field, for a wire $L={2000}~\mathrm{nm}$.
For $b<\Delta$, the wire is in the TTP and the Majorana mass is $\mathpzc{m}(x)>0$ along the whole wire.
For $\Delta<b<\sqrt{\delta V^2+\Delta^2}$, the wire is in the TIP with the Majorana mass $\mathpzc{m}(x)<0$ in the left section and $\mathpzc{m}(x)>0$ in the right section of wire
with nodes of the Majorana mass 
at $x=\pm \sqrt{b^2 - \Delta^2} L/\delta V$, i.e.,
the solutions of the equation $\sqrt{(\delta V x/L)^2+\Delta^2}\equiv b$.
For $b>\sqrt{\delta V^2+\Delta^2}$, the wire reaches the TNP with $\mathpzc{m}(x)<0$ along the whole wire.
The spatial dependence of the chemical potential in \cref{eq:potentialslope} and the resulting Majorana mass for different choices of the magnetic field are shown in \cref{fig:shapes}(c).

\begin{figure*}[t]
\centering
\includegraphics[width=\textwidth]{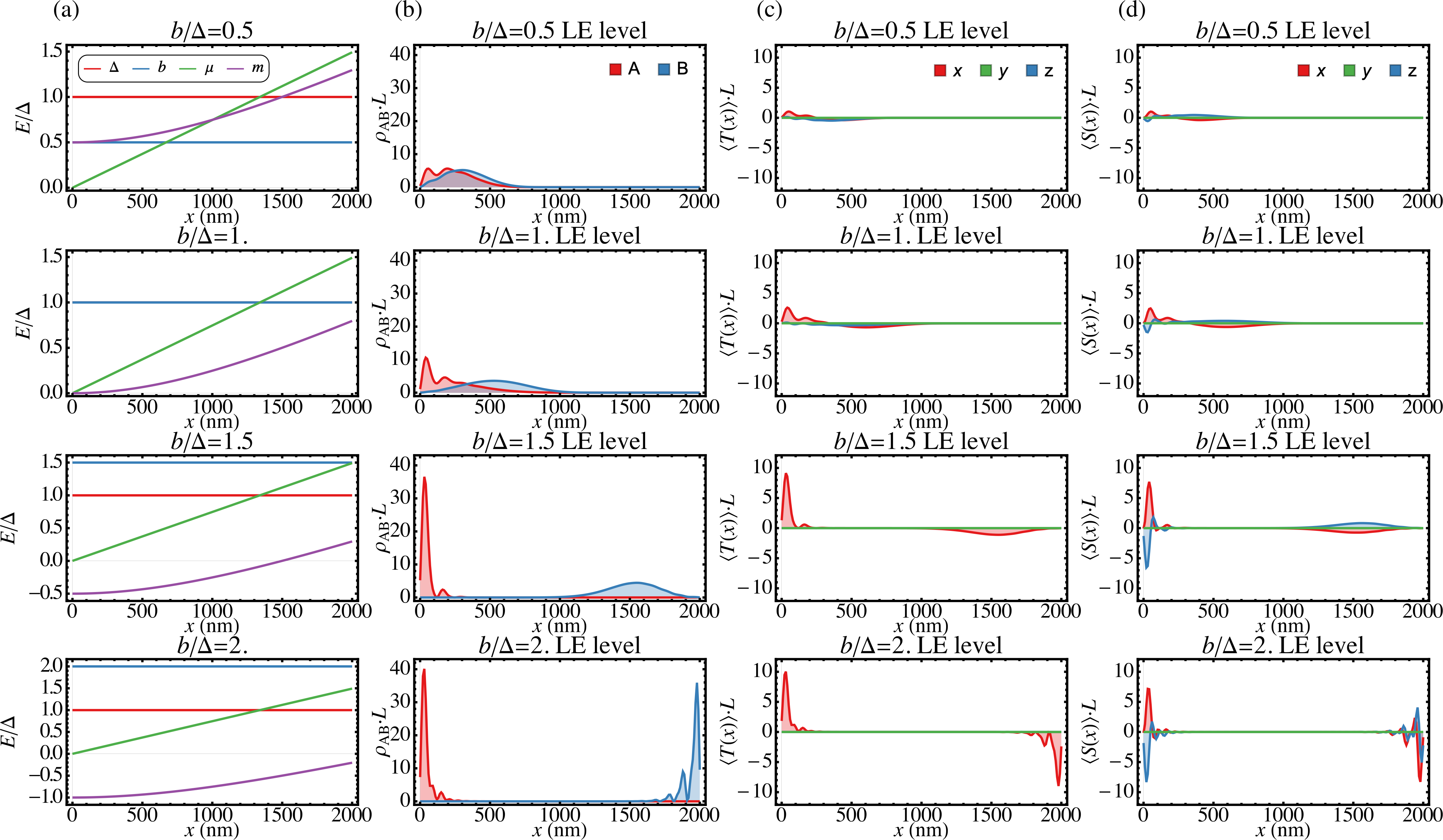} 
\caption{
Snapshots of the numerical results of \cref{fig:slopea} at several different Zeeman fields.
}
\label{fig:slopeb}
\end{figure*}

\Cref{fig:slopea} shows the energy spectra, the 
mutual overlap, transition probability, quasiparticle densities, expectation values of the Majorana pseudospin and spin of the MM, calculated for a Majorana wire with OBC and linear slope potential as a function of the Zeeman field.
\Cref{fig:slopeb} shows instead snapshots for different values of the Zeeman field.
As before, the wire goes from the TTP to the TNP passing by the TIP [shaded areas of \cref{fig:slopea}(a) and (b)] as the Zeeman field increases.
Again, a fo-MM detaches from the bulk excitation spectra in TTP and transmutes into 
a ps-MM at low energy localized near the left edges of the wire:
One of the MM become exponentially localized at the left edge as soon as the wire reaches the TIP, whereas a second MM becomes smoothly localized (i.e., with a Gaussian-shape peak) at the node of the local Majorana mass.
By increasing the Zeeman field, this second, rightmost MM moves continuously from the left to the right edges, following the node of the Majorana mass in the TIP\@.
Finally, when the wire reaches the TNP, this rightmost MM becomes exponentially localized at the right edge of the wire: the ps-MM is now maximally separated at the two opposite ends of the wire.
In this case, we did not calculate the fermion parity of the PBC groundstate, since the linear slope breaks the inversion symmetry between the two edges of the wire and therefore there is no obvious way to impose PBC\@.
As before, the Majorana pseudospin and spin increase,
whereas 
the mutual overlap, transition probability 
decays 
from $\Omega,W/\Delta E\approx1$ in the TTP 
reaching $\Omega,W/\Delta E\approx0$ in the TIP at higher Zeeman fields.
 
In the present case, the peaks of the quasiparticle density localized at the nodes of the Majorana mass are very broad if compared with the results obtained for the smooth potential barriers in the previous section.
This is justified by the fact that the potential slope is now linear, and does not exhibit any step-like features.
This is compatible with the results of the previous section, where slower variations of the Majorana mass lead to a broader localization peak of the ps-MM.

Moreover, we notice that the bulk gap does not close for any value of the Zeeman field, and the ps-MM simply detaches from the bulk spectra and become gradually pinned at zero energy as the Zeeman field increases.

\subsection{Crossover from impurity-induced ABS to MBS via smoothly interpolating potential}

We consider a potential given by the combination of a constant term and a spatially-varying term, as
\begin{equation}
\mu(x)=\mu-V(x),
\end{equation}
with $\mu=1.5\Delta$ and 
\begin{equation}\label{eq:smoothinterpolation}
V(x)={\delta V}
\left(S\left(\tfrac{2x-L+d}{2w}+\tfrac12\right)-S\left(\tfrac{2x-L-d}{2w}+\tfrac12\right)\right),
\end{equation}
where $S(x)$ is the cubic Hermite interpolator defined by
\begin{equation}\label{eq:interpolation}
S(x)=
\begin{cases}
0 & x<0\\
3x^2-2x^3 & 0\le x\le1\\
1 & x>1\\
\end{cases}
\end{equation}
The potential $V(x)$ is the sum of two smooth step function at a distance $d$, and interpolates between a localized impurity potential for $d=0$, $w\to 0$ and a hard wall potential $V(x)\to\delta V(\Theta(x)+\Theta(L-x))$ for $d=L$, $w\to 0$.
To obtain a smooth and continuous crossover between these two opposite limits, we vary the parameter $\delta V,w,d$ as a function of a control parameter $r$ as
\begin{equation}\label{eq:interpolation}
(\delta V,w,d)=
\begin{cases}
\left(r \mu,0,0\right)& \, 0\le r\le1,\\[5mm]
\left(\mu, (r-1) \frac{3L}{8},	(r-1) \frac{3L}{8}			 			\right) 	& \, 1\le r\le2,\\[5mm]
\left(\mu, \frac{3L}{8},		(r-2) \frac{L}4 + \frac{3L}8			\right) 	& \, 2\le r\le3,\\[5mm]
\left(\mu, (4-r)\frac{3L}{8},	(r-3) \frac{3L}{8} +\frac{5L}{8}		\right)		& \, 3\le r\le4.\\
\end{cases}
\end{equation}
The evolution of the potential $V(x)$ as a function of the control parameter $r$ and the resulting Majorana mass for $b=1.5\Delta$ are shown in \cref{fig:shapes}(d).
For $0< r\le 1$, the potential $V(x)$ is zero with the exception of $x=L/2$ in the middle of the wire, with an increasing peak height $V(L/2)=\delta V=r\mu$.
This regime models a localized impurity in the centre of the wire.
For $1\le r\le 2$, the impurity potential transmutes into a smooth bell-shaped potential.
For $2\le r\le3$, the bell-shaped potential separates into two smooth potential steps at a distance $d$, gradually reaching the edges of the wire.
This regime correspond to smooth inhomogeneous fields at the opposite edges of the wire.
For $3\le r\le4$, the two potential steps at the edges of the wire transition from a smooth potential step into a sharp hard wall potential and becomes constant $V(x)=\delta V$ at $r=4$.
This final regime correspond to a pristine nanowire in the topologically nontrivial phase. 
The potential $V(x)$ smoothly interpolates between an impurity-like potential ($r=0$) to the case of a perfectly uniform wire with open boundary conditions in the TNP ($r=4$) passing through a regime where the potential exhibits smooth spatial variations along the wire.
The choice of the interpolation path in \cref{eq:interpolation} is somewhat arbitrary and not unique, but serves the purpose: Showing the existence of a crossover between 
impurity-induced ABS and MBS, which can be described 
as the crossover between fo-MM and fs-MM.

\begin{figure*}[t]
\centering
\includegraphics[width=\textwidth]{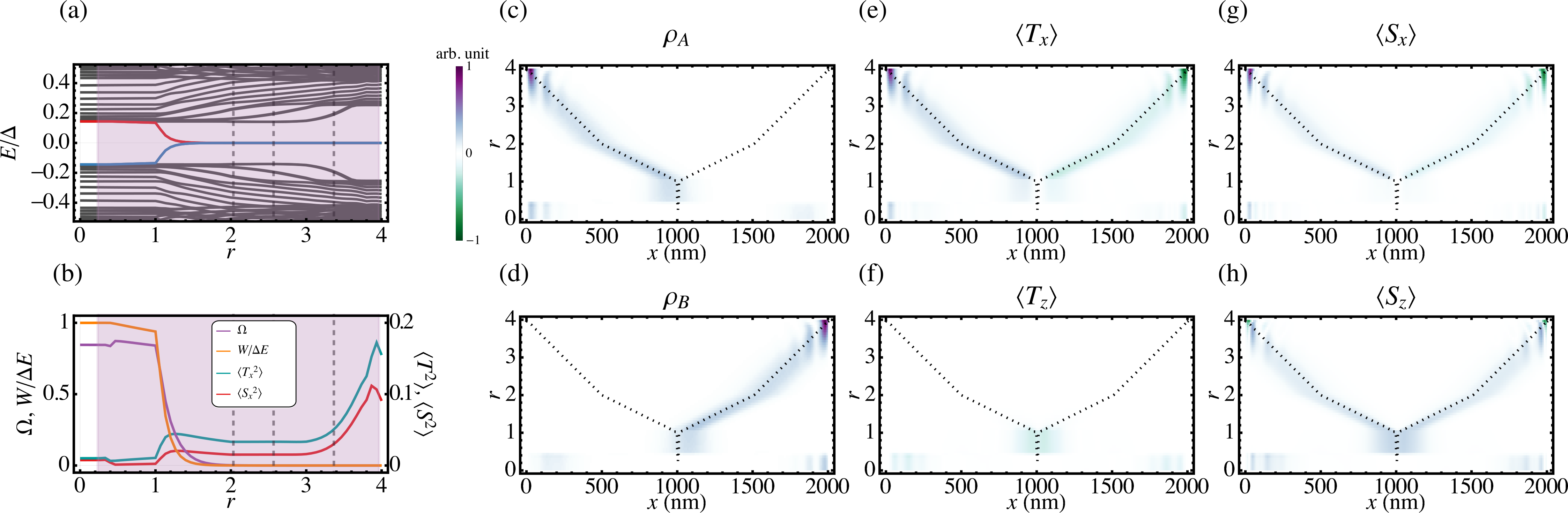} 
\caption{
Numerical results for a Majorana wire with OBC and smoothly interpolating potential as a function of the interpolating parameter $r$.
}
\label{fig:crossoveraOBC}
\end{figure*}

\begin{figure*}[t]
\centering
\includegraphics[width=\textwidth]{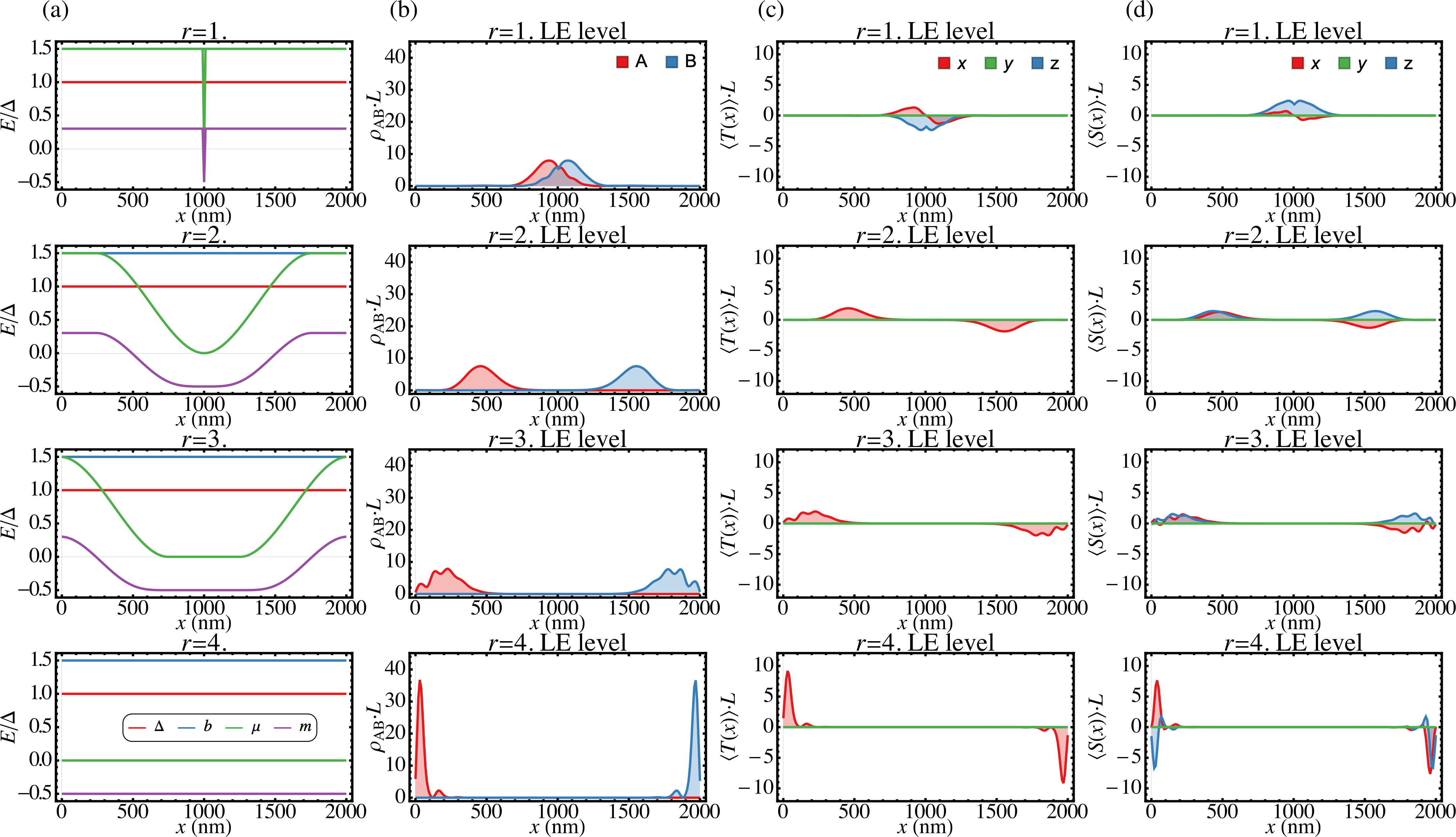} 
\caption{
Snapshots of the numerical results of \cref{fig:crossoveraOBC} at several different values of the interpolating parameter $r$.
}
\label{fig:crossoverbOBC}
\end{figure*}

\Cref{fig:crossoveraOBC} shows the energy spectra, mutual overlap, transition probability, quasiparticle densities, expectation values of the Majorana pseudospin and spin of the MM, calculated for a Majorana wire with OBC with Zeeman field $b=1.5\Delta$ and $L={2000}~\mathrm{nm}$
as a function of the interpolation parameter $r$.
\Cref{fig:crossoverbOBC} shows snapshots for different values of the control parameter $r$.
For small values of the control parameter $r\approx0$ such that $b<\sqrt{(\mu-\delta V)^2+\Delta^2}$,
the wire is in the TTP and the Majorana mass is $\mathpzc{m}(x)>0$ along the whole wire.
Increasing $0\le r\le 1$, the potential at $x=L/2$ increases and the central point assumes Majorana mass $\mathpzc{m}(L/2)<0$ when $b>\sqrt{(\mu-\delta V)^2+\Delta^2}$ and the wire enters the TIP [shaded areas of \cref{fig:crossoveraOBC}(a) and (b)].
Further increasing $1\le r\le 4$, the wire stays in the TIP with the Majorana mass $\mathpzc{m}(x)<0$ in the central section of the wire and $\mathpzc{m}(x)>0$ near the edges, with the nodes of the Majorana mass given by the solutions of the equation $\sqrt{V(x)^2+\Delta^2}\equiv b$.
The two MM move continuously from the central region of the wire to the edges, following the nodes of the Majorana mass in the TIP\@.
The whole crossover corresponds to a impurity-induced fo-MM for $r\le1$ which gradually transmutes into 
a inhomogeneities-induced ps-MM at low energy for $r\ge1$, moving from the center of the wire towards the edges, and smoothly localized (i.e., with a Gaussian-shape peak) at the nodes of the Majorana mass.
For $r=4$, the wire reaches the TNP and the Majorana mass is $\mathpzc{m}(x)<0$ along the whole wire, and the two MM become exponentially localized at the edges.
Notice that slow variations of the potential lead to a broader localization peak (i.e., Gaussian-like) of the ps-MM, whereas hard-wall potential barriers at $r=4$ lead to the exponential localization of the MM peaks.
As before, the mutual overlap, transition probability decays from $\Omega,W/\Delta E\approx1$ reaching $\Omega,W/\Delta E\approx0$ in the TIP at higher fields.
Conversely, the Majorana pseudospin and spin show a nearly constant value in the TIP, followed by a fast step-like increase when the MM come closer to the wire edges.

\begin{figure*}[t]
\centering
\includegraphics[width=\textwidth]{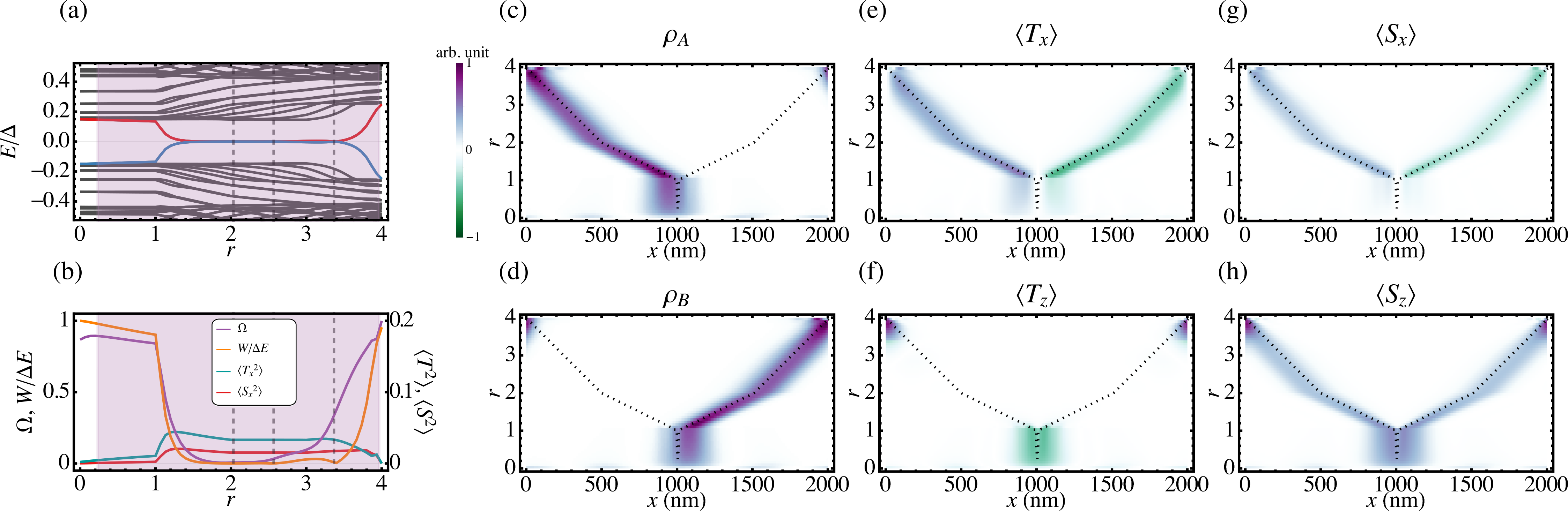} 
\caption{
Numerical results as in \cref{fig:crossoveraOBC} but for a Majorana wire with PBC\@.
}
\label{fig:crossoveraPBC}
\end{figure*}

The striking feature of the crossover analyzed here is that the bulk gap remains wide open nearly constant during the transition from the TIP into the TNP\@.
This is also true for the energy spectra calculated with PBC\@.
However, the fermion parity of the PBC groundstate shows three parity crossings which amount to the transition from $\mathcal P=1$ at $r\lesssim2
$ to $\mathcal P=-1$ for $r\gtrsim3$.
To understand better what happens, we show in \cref{fig:crossoveraPBC} the numerical results calculated for a Majorana wire with PBC as a function of the interpolation parameter $r$.
The PBC energy spectra in the TIP look very similar to the OBC spectra.
Notice that even with PBC, there is a ps-MM below the gap, which is localized at the nodes of the Majorana mass in the TIP\@.
This ps-MM is responsible for the fermion parity crossing and eventually, of the transition into the TNP\@.
Thus, the transition to the TNP is not accompanied by the closing and reopening of the bulk gap, but only to a fermion parity transition of a single subgap state.
Notice that, when the transition into the TNP is completed, the ps-MM transmutes into a bulk excitation, and there are no localized modes at the edges (as expected for PBC).
This corresponds to the fact that
the mutual overlap and transition probability 
decays 
from $\Omega,W/\Delta E\approx1$ 
reaching $\Omega,W/\Delta E\approx0$ in the TIP and come back to $\Omega,W/\Delta E\approx1$ in the TNP\@.
Conversely, the Majorana pseudospin and spin show a nearly constant plateau in the TIP, and consequently vanishes in the TNP\@.

\section{Discussion}

We characterized the different physical regimes in terms of topology and localization properties of the subgap states.
To characterize topology, 
we defined and calculated both the global and local topological invariants in the presence of inhomogeneities and impurities, and show how these two quantities differ from the topological invariant of a uniform wire.
In particular, three different phases appear:
the homogeneous topologically trivial phase	(TTP),
the homogeneous topologically nontrivial phase (TNP),
and a topologically inhomogeneous phase (TIP) separating the first two phases.
In the two homogenous phases, the local topological invariant is uniform along the wire, and coincides with the global topological invariant, being either trivial (TTP) or nontrivial (TNP).
In the TIP, the local topological invariant is nonuniform along the wire, Majorana modes localize near the points where the local topological invariant changes sign, i.e., at the nodes of the local Majorana mass $\mathpzc{m}(x)$.
Since the fermion parity of the PBC groundstate is even and odd respectively in the TTP and in the TNP, the subgap levels must exhibit an odd number of fermion parity crossing in the TIP\@.
However, the topological transition does not necessarily correspond to the condition $|b|\equiv|b_c|=\sqrt{\mu^2+\Delta^2}$, and does not necessarily coincide with the closing and reopening of the bulk gap, but only to a parity crossing of the LE subgap state.

To characterize the localization properties of the subgap state, we define and calculate the wavefunctions, quasiparticle densities, Majorana pseudospin and spin polarizations, 
density overlaps and transition probabilities between the Majorana components.
This lead us to distinguish between fully-separated Majorana modes (fs-MM) with $\Omega=W=0$, partially-separated Majorana modes (ps-MM) with $0<\Omega<1$ and $W>0$, and fully-overlapping Majorana modes (fo-MM) with $\Omega=1$ and $W\approx\Delta E$.
Notice that these definitions are not based on topology, but only on the intrinsic properties of the subgap wavefunctions, and well-defined both for infinite and finite wires.
In this context, topologically protected MBS can be unambiguously defined as a couple of fs-MM localized at the opposite ends of an infinite wire in the homogeneous topologically nontrivial phase, or in the inhomogeneous phase with nontrivial global topological invariant, as long as the nodes of the Majorana mass are restricted in a finite segment at a finite distance from the wire edges.
Thus, strictly speaking fs-MM are a limiting case which only exists in infinite-size systems.
In finite-size wires, only ps-MM and fo-MM can exist.

Any occurrence of Majorana/Andreev crossover can be described as a transition along the continuous interval $\Omega\in[0,1]$ between the two limiting cases of fs-MM ($\Omega=0$) and fo-MM ($\Omega=1$).
This crossover can be described as the fusion of two Majorana modes (fs-MM, $\Omega=0$) localized at the nodes of the Majorana mass $\mathpzc{m}(x)\equiv0$ (or at the edges of the wire)
into a single Dirac fermion (fo-MM, $\Omega=1$) when the two nodes become close together, merge, and disappear.
The crossover from impurity-induced ABS to quasi-MBS~\cite{pan_crossover_2021} coincides with the crossover
from a single fo-MM ($\Omega=1$) localized around the impurity in the TTP, to a ps-MM ($0<\Omega<1$) in the TIP\@.
On the other hand, the crossover from inhomogeneities-induced ABS (quasi-MBS) to MBS~\cite{moore_two-terminal_2018,stanescu_robust_2019,avila_non-hermitian_2019} coincides with the crossover
from a ps-MM ($0<\Omega<1$) in the TIP, to a fs-MM ($\Omega=1$) in the TNP\@.
Finally, there exists a crossover from impurity-induced ABS to topologically-protected MBS, i.e., 
from a single fo-MM ($\Omega=1$) localized around the impurity in a TTP, to a fs-MM ($\Omega=0$) localized at the wire ends in the TNP\@.

The presence of a continuous crossover between trivial and nontrivial subgap modes without a TQPT suggests that it may be not physically possible to univocally and unambiguously distinguish between ``true'' MBS, and other varieties of subgap modes.
In fact, any subgap mode is characterized by the mutual overlap $X$ of its Majorana components, on a continuous crossover $X\in[0,1]$ of ps-MM, with the limiting cases of fo-MM with overlap $X=1$ and fs-MM with $X=0$.
Pragmatically, however, it is still relevant to determine the extent of the spatial separation of the MM, i.e., how well separated are the Majorana components $\gamma_A$ and $\gamma_B$ along the wire, where are these components localized, and how robust they are against disorder.
The degree of spatial separation between MM and their localization with respect to the wire edges can be determined by nonlocal experiments~\cite{zocher_modulation_2013,liu_majorana_2013,li_probing_2014,stanescu_nonlocality_2014,haim_current_2015,haim_signatures_2015,clarke_experimentally_2017,schuray_fano_2017,prada_measuring_2017,moore_two-terminal_2018,rosdahl_andreev_2018,liu_distinguishing_2018,hell_distinguishing_2018,ricco_spin-dependent_2019,lai_presence_2019,vuik_reproducing_2019,awoga_supercurrent_2019,mishmash_dephasing_2020,pan_physical_2020,pan_three-terminal_2021,cayao_distinguishing_2021}.

For brevity, we considered in this work only configurations with only two MM localized inside the wire, corresponding to the presence of only two nodes of the Majorana mass $\mathpzc{m}(x)$, and we only considered spatial variations of the chemical potential.
The results presented here can be easily generalized to cases where both the chemical potential, Zeeman field, and superconducting pairing are spatially modulated along the wire, and to configurations where the Majorana mass exhibit more than two nodes within the wire.

The description of subgap states as partially separated modes localized at the nodes of the Majorana mass is well-suited only for inhomogeneous fields having variations on length scales comparable with the Majorana localization length, and such that the nodes of the Majorana mass correspond to the presence of a few, well-separated subgap modes.
This description breaks down, e.g., in the regime of strong disorder~\cite{das-sarma_disorder-induced_2021,pan_generic_2020}. 
In this case, the Majorana mass may exhibit large oscillations and its nodes may become extremely dense and close to each other, which results in the proliferation of subgap modes hybridized over the whole length of the wire.
In this regime, it is not meaningful to model the Majorana nanowire via an effective low-energy Hamiltonian describing the hybridization of a few subgap Majorana modes, as in \cref{eq:Heff}.
The transition between impurity-induced ABS described here and strong disorder is left for future work.
Moreover, at very short lengths, the discreteness of the atomic lattice cannot be neglected.
This may lead to a quasiperiodic regime where the competition between lattice and field length scales give rise to fractal energy bands (Hofstadter butterfly) and Anderson localization~\cite{cai_topological_2013,degottardi_majorana_2013a,degottardi_majorana_2013b}, or to the presence of nontrivial ABS~\cite{marra_topologically_2019}.
Another example of ABS which cannot be described as separated Majorana modes localized at the nodes of the Majorana mass is the intrinsic ABS described in Ref.~\cite{huang_metamorphosis_2018}.

\section{Conclusions}

In summary, we characterized the continuous crossover between Andreev and Majorana bound states in terms of their Majorana components, showing how the crossover between impurity-induced to inhomogeneities-induced subgap modes and to Majorana bound states can be described as the crossover between fully-overlapping to partially-separated and fully-separated Majorana modes.
The localization of the Majorana modes have been characterized via their quasiparticle density, mutual overlap, transition probability, Majorana pseudospin, and spin polarizations along the wire.
We showed that these partially-separated modes localize at the nodes of the Majorana mass, being either exponentially or smoothly localized respectively in the case of sharp and smooth variations of the Majorana mass near the node.
In particular, Majorana bound states localized at the wire edges correspond to the limiting case of fully-separated Majorana modes realized when the distance between the Majorana modes is infinite.
We discussed the presence of a topological inhomogeneous phase which is intermediate between the topologically trivial and nontrivial phases, and is characterized by a local topological invariant spatially-varying along the wire.
We evidenced the absence of a global topological phase transition:
Different regimes are still characterized by a different fermion parity.
However, changes of the fermion parity do not correspond necessarily to the closing of the particle-hole gap, but may occur via the parity crossing of the lowest energy subgap state.
This work suggests that there may be no sharp distinction between ``true'' Majorana bound states and other varieties of zero-energy or near-zero-energy subgap modes in the presence of spatially-varying potentials and impurities.
Indeed, subgap modes in Majorana wires interpolate continuously between fully-overlapping, partially-separated, and fully-separated Majorana modes.
 
\begin{acknowledgments}
The work of P.~M.~is supported by the Japan Science and Technology Agency (JST) of the Ministry of Education, Culture, Sports, Science and Technology (MEXT), JST CREST Grant.~No.~JPMJCR19T2, and by the Japan Society for the Promotion of Science (JSPS) Grant-in-Aid for Early-Career Scientists (Grant No.~20K14375).
\end{acknowledgments}

\section*{Data availability statement}
The code used for the numerical calculations and the resulting data can be found on Zenodo~\cite{marra_data_2021}.

\end{document}